\documentclass[a4paper,USenglish]{lipics}

\usepackage{algorithm,booktabs}
\usepackage{algpseudocode}
\usepackage{tikz}
\usepackage{pgfplots}
\usepackage{mathtools}
\usetikzlibrary{3d,calc,patterns,shadings}
\definecolor{darkgreen}{rgb}{0,0.66,0}

\renewcommand{\vec}[1]{\mathbf{#1}}

\newcommand{\eps}{\varepsilon}
\newcommand{\vc}[1]{\boldsymbol{#1}}
\renewcommand{\vec}[1]{\boldsymbol{#1}}

\newcommand{\cC}{\mathcal{C}}
\newcommand{\cD}{\mathcal{D}}

\newcommand{\cH}{\mathcal{H}}

\newcommand{\cL}{\mathcal{L}}

\newcommand{\cP}{\mathcal{P}}

\newcommand{\cS}{\mathcal{S}}
\newcommand{\cT}{\mathcal{T}}

\newcommand{\cW}{\mathcal{W}}

\newcommand{\R}{\mathbb{R}}

\newcommand{\ip}[2]{\left\langle{#1},{#2}\right\rangle}
\newcommand{\rhoq}{\rho_{\mathrm{q}}}

\newcommand{\rhou}{\rho_{\mathrm{u}}}
\newcommand{\alphaq}{\alpha_{\mathrm{q}}}
\newcommand{\alphau}{\alpha_{\mathrm{u}}}

\title{Faster tuple lattice sieving using spherical locality-sensitive filters\footnote{This paper merges earlier results from~\cite{laarhoven15nns} regarding \textit{dense} data sets (the results for \textit{sparse} data sets were previously merged into~\cite{andoni17}) with new results about tuple lattice sieving.}}

\titlerunning{Faster tuple lattice sieving using spherical LSF}

\author[1]{Thijs Laarhoven}
\affil[1]{IBM Research\\
  R\"{u}schlikon, Switzerland\\
  \texttt{mail@thijs.com}}
\authorrunning{Thijs Laarhoven} 
\Copyright{Thijs Laarhoven}

\subjclass{F.2 Analysis of algorithms and problem complexity}

\keywords{near neighbor searching, locality-sensitive hashing/filtering (LSH/LSF), lattice algorithms, shortest vector problem (SVP), lattice sieving}

\serieslogo{}
\volumeinfo
  {Billy Editor and Bill Editors}
  {2}
  {Conference title on which this volume is based on}
  {1}
  {1}
  {1}
\EventShortName{}
\DOI{10.4230/LIPIcs.xxx.yyy.p}

\begin{document}

\maketitle

\begin{abstract}
To overcome the large memory requirement of classical lattice sieving algorithms for solving hard lattice problems, Bai--Laarhoven--Stehl\'{e} [ANTS 2016] studied tuple lattice sieving, where tuples instead of pairs of lattice vectors are combined to form shorter vectors. Herold--Kirshanova [PKC 2017] recently improved upon their results for arbitrary tuple sizes, for example showing that a triple sieve can solve the shortest vector problem (SVP) in dimension $d$ in time $2^{0.3717d + o(d)}$, using a technique similar to locality-sensitive hashing for finding nearest neighbors.

In this work, we generalize the spherical locality-sensitive filters of Becker--Ducas--Gama--Laarhoven [SODA 2016] to obtain space-time tradeoffs for near neighbor searching on dense data sets, and we apply these techniques to tuple lattice sieving to obtain even better time complexities. For instance, our triple sieve heuristically solves SVP in time $2^{0.3588d + o(d)}$. For practical sieves based on Micciancio--Voulgaris' GaussSieve [SODA 2010], this shows that a triple sieve uses less space \textit{and} less time than the current best near-linear space double sieve. 
\end{abstract}


\section{Introduction}

\subparagraph{Lattice-based cryptography.} Over the past few decades, lattice-based cryptography has emerged as a strong candidate for constructing efficient cryptographic primitives~\cite{micciancio02, micciancio09, peikert16}. Its security is based on the hardness of hard lattice problems such as the shortest vector problem (SVP): given a description of a lattice, find a shortest non-zero lattice vector. To accurately choose parameters for lattice-based cryptographic primitives, it is crucial to be able to accurately estimate the computational complexity of solving these problems.

\subparagraph{SVP algorithms.} Currently there are two main classes of practical algorithms for solving SVP in high dimensions. Lattice enumeration~\cite{kannan83, fincke85, gama10, micciancio15, aono17} uses superexponential time and polynomial space in the lattice dimension $d$ to solve SVP, while lattice sieving~\cite{ajtai01, nguyen08, micciancio10b, laarhoven15crypto, becker16lsf} requires time and memory both single exponential in $d$. In high dimensions sieving will clearly be faster than enumeration, but the large memory requirement both limits our ability to execute sieving algorithms in high dimensions, and it significantly slows down sieving in practice due to the costly memory accesses~\cite{mariano15, mariano16pdp, mariano17}.

\subparagraph{Tuple lattice sieving.} Tuple lattice sieving aims to overcome the main drawback of classical lattice sieving methods by using less memory at the cost of more time, offering a tradeoff between sieving and enumeration. After Bai--Laarhoven--Stehl\'{e}~\cite{bai16} made a first step towards analyzing tuple lattice sieving, Herold--Kirshanova~\cite{herold17} significantly improved upon this by (i) proving what are the memory requirements for arbitrary tuple sizes; (ii) analyzing which configurations of tuples one should look for; (iii) showing that tuple sieving can be modified to use much less time; and (iv) showing how a near neighbor-like method called Configuration Extension can be used to further reduce the asymptotic time complexity. As an example, their optimized triple sieve requires $2^{0.1887d + o(d)}$ memory and $2^{0.3717d + o(d)}$ time.

\subparagraph{Locality-sensitive filtering.} For finding near neighbors in the context of lattice sieving, Becker--Ducas--Gama--Laarhoven~\cite{becker16lsf} introduced spherical locality-sensitive filters (LSF), to date achieving the best heuristic time complexities for sieving with pairwise reductions: using either $2^{0.2925d + o(d)}$ time and memory, or using $2^{0.3684d + o(d)}$ time with only $2^{0.2075d + o(d)}$ memory.\footnote{Theoretically, one can achieve $2^{0.2925d + o(d)}$ time using only $2^{0.2075d + o(d)}$ memory using a sieve based on the less practical Nguyen--Vidick's sieve~\cite{nguyen08}. In practice no one has ever used the Nguyen--Vidick sieve to solve SVP in high dimensions, due to the large hidden order terms~\cite{svp}.} The original LSF framework was only described to provide a balanced tradeoff between the time and space complexities, and after an early version of this work\footnote{Historically, in late 2015 the preprint~\cite{laarhoven15nns} first described these space-time tradeoffs. For the 2017 paper~\cite{andoni17}, the tradeoffs for sparse data sets of~\cite{laarhoven15nns} were then merged with the lower bounds of~\cite{andoni16}, whereas~\cite{christiani17} independently obtained similar lower bounds for sparse regimes in 2017 as well. The present paper merges results from~\cite{laarhoven15nns} for dense data sets with new results for tuple lattice sieving.} showed how to obtain arbitrary space-time tradeoffs for the angular distance, both Christiani~\cite{christiani17} and Andoni--Laarhoven--Razenshteyn--Waingarten~\cite{andoni17} showed that these tradeoffs can be extended to any $\ell_p$ distance ($p \in [1, 2]$), and that the resulting tradeoffs are in fact provably optimal for sparse data sets of size $n = 2^{o(d)}$. 

\subsubsection*{Contributions} 

Our contributions are essentially three-fold:
\begin{itemize}
\item We generalize the spherical locality-sensitive filters from~\cite{becker16lsf} to obtain explicit space-time tradeoffs for uniformly random data sets on the sphere of size $n = 2^{\Theta(d)}$.
\item We adapt tuple lattice sieving to include arbitrary near neighbor techniques, using a new transform to guarantee that each search is done over uniformly random data sets.
\item We finally apply the space-time tradeoffs to tuple lattice sieving to obtain improved asymptotics for the time complexities of tuple lattice sieving for arbitrary tuple sizes.
\end{itemize}
For the triple sieve, this leads to a time complexity for solving SVP of $2^{0.3588d + o(d)}$, improving upon the $2^{0.4812d + o(d)}$ of~\cite{bai16} and the $2^{0.3717d + o(d)}$ of~\cite{herold17}, while maintaining a space complexity of $2^{0.1887d + o(d)}$. Using the same amount of memory as a double sieve ($2^{0.2075d + o(d)}$), our triple sieve can solve SVP in time $2^{0.3317d + o(d)}$. These complexities hold for triple sieves based on Micciancio--Voulgaris' GaussSieve~\cite{micciancio10b} as well, which means that compared to the best GaussSieve-based near-linear space double sieve of~\cite{becker16lsf}, running in time $2^{0.3684d + o(d)}$ and space $2^{0.2075d + o(d)}$, we can either save significantly on the time complexity, or save both on the time \textit{and} on the required amount of memory.\footnote{Here, ``near-linear space'' means that we consider GaussSieve-based sieves with memory bounded by the list size, $2^{0.2075d + o(d)}$. Using more memory one can solve SVP in time and space $2^{0.2925d + o(d)}$~\cite{becker16lsf}.} This is a rather surprising result, since previous results suggested that tuple lattice sieving only offers a tradeoff of using less space at the cost of more time.



\section{Preliminaries}
\label{sec:prelim}

\subparagraph*{Notation.} Throughout, $d$ denotes the dimensionality of the space, $n$ commonly denotes the (exponential) size of a list of vectors in $\mathbb{R}^d$, and $k$ denotes the tuple size. We write vectors in boldface (e.g.\ $\vc{x}$), and denote their Euclidean norms by $\|\vc{x}\|$. The unit sphere in $d$ dimensions is denoted by $\mathcal{S}^{d-1}$. Given two vectors $\vc{x}, \vc{y} \in \mathbb{R}^d$, we denote their inner product by $\langle \vc{x}, \vc{y} \rangle$, and we write $\phi(\vc{x}, \vc{y}) := \arccos \frac{\langle \vc{x}, \vc{y} \rangle}{\|\vc{x}\| \cdot \|\vc{y}\|}$ for their common angle. We write $\mu$ for the canonical Lebesgue measure over $\R^d$, and we denote half-spaces by $\cH_{\vec u, \alpha} := \{\vec x \in \R^d: \ip{\vec u}{\vec x} \geq \alpha\}$.

\subparagraph*{Subsets of the unit sphere.} We recall some properties about geometric objects on the unit sphere, similar to~\cite{becker16lsf}. For constants $\alpha_1, \alpha_2 \in (0,1)$ and vectors $\vec u_1, \vec u_2 \in \cS^{d-1}$ we denote spherical caps and wedges (intersections of spherical caps) by $\cC_{\vec u_1, \alpha_1} := \cS^{d-1} \cap \cH_{\vec u_1, \alpha_1}$ and $\cW_{\vec u_1, \alpha_1, \vec u_2, \alpha_2} := \cS^{d-1} \cap \cH_{\vec u_1,\alpha_1} \cap \cH_{\vec u_2, \alpha_2} = \cC_{\vec u_1, \alpha_1} \cap \cC_{\vec u_2, \alpha_2}$. Denoting $\theta = \phi(\vc{u}_1, \vc{u}_2)$, the relative volumes of these objects can be estimated as follows \cite[Lemmata 2.1 and 2.2]{becker16lsf}: 
\begin{align}
\cC(\alpha_1) &:= \frac{\mu(\cC_{\vec u_1,\alpha_1})}{\mu(\cS^{d-1})} = d^{\Theta(1)} \left(1 - \alpha_1^2\right)^{d/2}, \\
\cW(\alpha_1,\alpha_2,\theta) &:= \frac{\mu(\cW_{\vec u_1, \alpha_1, \vec u_2, \alpha_2})}{\mu(\cS^{d-1})} = d^{\Theta(1)} \left(1 - \frac{\alpha_1^2 + \alpha_2^2 - 2 \alpha_1 \alpha_2 \cos \theta}{\sin^2 \theta}\right)^{d/2}.
\end{align}
 
\subparagraph*{Lattices.} Lattices are discrete subgroups of $\mathbb{R}^d$. Given a basis $B = \{\vc{b}_1, \dots, \vc{b}_d\} \subset \R^d$ of linearly independent vectors, the lattice generated by $B$, denoted $\cL$ or $\cL(B)$, is given by $\cL(B) := \{\sum_{i=1}^d \lambda_i \vc{b}_i : \lambda_i \in \mathbb{Z}\}$. The shortest vector problem (SVP) is: given a description of a lattice (e.g.\ a basis of the lattice), find a shortest non-zero vector in this lattice. 

\subparagraph{Lattice sieving.} The fastest (heuristic) algorithms for solving SVP in high dimensions are based on lattice sieving, originally described in~\cite{ajtai01} and later improved in e.g.~\cite{nguyen08, pujol09, micciancio10b, wang11, zhang13, laarhoven15crypto, becker15nns, laarhoven15latincrypt, becker16cp, becker16lsf}. Given a basis of a lattice $\cL$, lattice sieving attempts to solve SVP by first generating a long list $L \subset \cL$ of relatively long lattice vectors, and then iteratively combining vectors in $L$ to form shorter and shorter lattice vectors until a shortest lattice vector is found. Key properties of lattices used in sieving are (1) if $\vc{x}_1, \vc{x}_2 \in \cL$, then also $\vc{x}_1 \pm \vc{x}_2 \in \cL$; and (2) if $\phi(\vc{x}_1, \vc{x}_2) < \frac{\pi}{3}$, then $\min\{\|\vc{x}_1 \pm \vc{x}_2\|\} < \max\{\|\vc{x}_1\|, \|\vc{x}_2\|\}$, so that adding/subtracting these vectors leads to a shorter lattice vector. After generating a long list $L$, classical sieving algorithms usually proceed by picking $\vc{x}_1 \in L$, searching for a vector $\vc{x}_2 \in L$ with angle less than $\frac{\pi}{3}$ with $\vc{x}_2$, performing a \textit{reduction} $\vc{x}_1 \gets \vc{x}_1 \pm \vc{x}_2$, and repeating. 

\subparagraph*{Sieving complexities.} Under the heuristic assumption that when normalized, vectors in $L$ are uniformly distributed on $\cS^{d-1}$, one can show that the list size under pairwise reductions scales as $|L| = \sin(\frac{\pi}{3})^{-d + o(d)} = 2^{0.2075d + o(d)}$~\cite{nguyen08}. A naive linear search thus leads to a quadratic time complexity of $2^{0.4150d + o(d)}$~\cite{nguyen08}, whereas the best near neighbor techniques can speed this up to $2^{0.3684d + o(d)}$ while retaining the same memory~\cite{becker16lsf}, and offer a tradeoff between using even more memory and less time. Minimizing the time complexity leads to time and space complexities both equal to $2^{0.2925d + o(d)}$~\cite{becker16lsf}.\footnote{Theoretically, one can achieve time complexity $2^{0.2925d + o(d)}$ with space complexity $2^{0.2075d + o(d)}$ using the NV-sieve of~\cite{nguyen08} as a starting point, instead of the more practical GaussSieve of~\cite{micciancio10b}. In practice no records for solving SVP in high dimensions have ever been obtained using the NV-sieve~\cite{svp}.} To date, all classical sieving algorithms using pairwise reductions seem bound by a minimum space complexity of $2^{0.2075d + o(d)}$, which puts a serious restriction on its practicality in high dimensions.
 
\subparagraph*{Tuple lattice sieving.} Tuple lattice sieving, originally proposed in~\cite{bai16} and later improved in~\cite{herold17}, aims to overcome this large space requirement by combining multiple list vectors for reductions: instead of looking for short vectors $\vc{x}_1 \pm \vc{x}_2$, one looks for short combinations $\vc{x}_1 \pm \dots \pm \vc{x}_k$ with all $\vc{x}_i \in L$.\footnote{As described in \cite[Section 5.2]{bai16}, for tuple sizes $k > 4$, one may want to look for combinations $\lambda_1 \vc{x}_1 + \dots + \lambda_k \vc{x}_k$ with $\lambda_i$ potentially coming from a slightly larger set than $\{-1, 1\}$.} By considering a larger number of more intricate combinations of the list vectors, one hopes to reduce the required list size to make progress. As conjectured in~\cite{bai16} and later proved in~\cite{herold17}, this is indeed the case:
\begin{lemma}[List sizes for tuple sieving {\cite[Theorem 3]{herold17}}] \label{lem:ls}
Let $L_k \subset \cS^{d-1}$ consist of $n$ uniformly random unit vectors. Let $L_k'$ consist of all $k$-tuples $(\vc{x}_1, \dots, \vc{x}_k) \in L_k^k$ satisfying $\|\vc{x}_1 \pm \dots \pm \vc{x}_k\| < 1$. Then $|L_k'| / |L_k| \geq 1 - o(1)$ as $d \to \infty$ iff:
\begin{align}
|L_k| \geq \left(\frac{k^{k/(k-1)}}{k+1}\right)^{d/2 + o(d)}.
\end{align} 
\end{lemma}
In other words, to make sure that (1) we make progress in finding shorter and shorter lattice vectors, and (2) the output list $L_k'$ is not much smaller than the input list $L_k$, the above lemma tells us how big $L_k$ needs to be. For $k = 2$ this gives us the classical bound $L_2 = 2^{0.2075d + o(d)}$ for sieving with pairwise reductions~\cite{nguyen08, micciancio10b, laarhoven15crypto, becker16lsf}.

Naively, finding good combinations of $k$ list vectors can be done by considering all $k$-tuples of vectors in $L_k$, requiring time $\tilde{O}(|L_k|^k)$. In~\cite{bai16} techniques were described to reduce the search space and find (almost) all good $k$-tuples faster, and~\cite{herold17} further improved upon this by stating exactly which \textit{configurations} of tuples one should be looking for. In particular, this reduces the ``global'' search condition on the entire $k$-tuple (the sum being short) to ``local'' conditions on pairs of vectors, greatly simplifying and speeding up the search procedure.
\begin{lemma}[Dominant configurations {\cite[Theorem 2]{herold17}}] \label{lem:conf}
Let $\eps > 0$, and let $L_k$ and $L_k'$ as in Lemma~\ref{lem:ls}. Let $L_k'' \subset L_k'$ denote the $k$-tuples in $L_k'$ with pairwise inner products satisfying:
\begin{align}
\left|\ip{\vc{x}_i}{\vc{x}_j} + \tfrac{1}{k}\right| \leq \eps, \qquad (i \neq j). 
\end{align}
Then $|L_k''|/|L_k'| = 1 - o(1)$ as $d \to \infty$.
\end{lemma}
In other words, for $k$-tuple sieving, it is sufficient to only look for tuples for which all pairwise inner products are essentially equal to $-\frac{1}{k}$, such that together with $\sum_{i=1}^d \vc{x}_i$ these vectors are the vertices of a simplex with the origin as its center~\cite{herold17}.


\subparagraph*{Finding near neighbors on the sphere.} To describe the near neighbor technique of~\cite{becker16lsf}, we first introduce the near neighbor problem on the sphere as follows. Note that we make explicit assumptions on the distribution of points in the data set, simplifying later analyses.
\begin{definition}[Near neighbor on the sphere]
Let $L$ consist of $n$ points drawn uniformly at random from $\cS^{d-1}$, and let $\theta \in (0, \frac{\pi}{2})$. The $\theta$-near neighbor problem is to preprocess $L$ such that, given a query vector $\vc{q} \in \cS^{d-1}$, one can quickly find a point $\vc{p} \in L$ with $\phi(\vc{p}, \vc{q}) \leq \theta$.
\end{definition}
Depending on the magnitude of $n$ we further make a distinction between the near neighbor problem for \textit{sparse} data sets ($n = 2^{o(d)}$) and for \textit{dense} data sets ($n = 2^{\Theta(d)}$). In many applications of near neighbor searching one is interested in sparse data sets, and various lower bounds matching upper bounds have been derived for this regime~\cite{odonnell14, andoni15cp, andoni17, christiani17}. In this paper we will focus on the dense regime, of interest in the application of lattice sieving.


\section{Spherical locality-sensitive filters for random dense data sets} 
\label{sec:dense}

\subparagraph*{Spherical locality-sensitive filters.} To solve the near neighbor problem on the sphere, Becker--Ducas--Gama--Laarhoven~\cite{becker16lsf} introduced spherical locality-sensitive filters, inspired by e.g.\ the spherical cap LSH of~\cite{andoni14}. The idea is to create a data structure of many \textit{filter} buckets, where a bucket contains vectors which are close to a randomly drawn filter vector $\vc{u} \in \cS^{d-1}$. Here, two vectors are considered close iff $\ip{\vc{u}}{\vc{p}} \geq \alpha$ (or equivalently $\phi(\vc{u}, \vc{p}) \leq \arccos \alpha$) for $\alpha \in (0,1)$ to be chosen later. Ideally one generates $f \gg 1$ of these buckets, each with $\vc{u} \in \cS^{d-1}$ chosen independently and uniformly at random. Inserting vectors into the data structure corresponds to finding the right filters for insertion, while queries first retrieve the filters that are close to the target, and then searches for near neighbors in those buckets.

\subparagraph*{Structured filters.} A naive implementation of this idea would lead to an impractically large overhead of finding the filters that a vector is in -- one would have to go through all $f$ filters one by one. To surmount this problem, a small amount of \textit{structure} is added to the filter vectors $\vc{u}$, making them dependent: small enough so that their joint distribution is sufficiently close to $f$ independent random vectors, but large enough to ensure that finding the filters that a vector is in can be done in time proportional to the number of filters that a vector is in, rather than proportional to the total number of filters $f$. This technique was later called ``tensoring'' in~\cite{christiani17}, and replaced with a tree-based data structure in~\cite{andoni17}. For further details regarding this technique we refer the reader to~\cite{becker16lsf} -- below, we will simply assume that filter vectors are essentially independent, and decoding can be done efficiently.

\subparagraph*{Cost analysis.} To obtain a tradeoff between the query and update costs of this data structure, we introduce two different parameters $\alphaq$ and $\alphau$. Each vector $\vc{p} \in L$ is inserted into all buckets with a filter vector $\vc{u}$ satisfying $\ip{\vc{u}}{\vc{p}} \geq \alphau$. Querying for near neighbors to $\vc{q}$ is done by retrieving all filters $\vc{u}$ with $\ip{\vc{u}}{\vc{q}} \geq \alphaq$, and going through all vectors in these buckets, looking for near neighbors. Some observations (see also Figure~\ref{fig:geometry}):
\begin{itemize}
\item \textbf{Updates:} A vector is added to a filter with probability $\cC(\alphau)$, and is therefore on average contained in $f \cdot \cC(\alphau)$ filters. In total, $|L| = n$ vectors are thus expected to generate $n \cdot f \cdot \cC(\alphau)$ bucket entries, or $n \cdot \cC(\alphau)$ vectors per bucket.
\item \textbf{Queries:} A vector queries a random filter with probability $\cC(\alphaq)$, and so on average a query returns $f \cdot \cC(\alphaq)$ filters, with $f \cdot \cC(\alphaq) \cdot n \cdot \cC(\alphau)$ collisions in the buckets.
\item \textbf{Collisions:} Two vectors at a specified angle $\theta$ are both contained in a random filter with probability $\cW(\alphaq, \alphau, \theta)$, and collide with constant probability iff $f = O(1 / \cW(\alphaq, \alphau, \theta))$.
\item \textbf{Space:} The dominant space requirements are having to store the $n$ vectors, and having to store the $n \cdot f \cdot \cC(\alphau)$ bucket entries in memory.
\end{itemize}
Constructing the entire data structure can be done by inserting each vector $\vc{p} \in L$ in the data structure one by one, which under the assumption that the time spent on finding relevant filters is proportional to the number of filters it should be inserted in, can be done in time proportional to the memory requirement $n \cdot f \cdot \cC(\alphau)$.

\begin{figure*}[!t]
\centering
\begin{tikzpicture}

\node at (-1cm, -1.5cm) {\Large $\cS^{d-1}$};

\begin{scope}
  \clip (1.25cm, -2.5cm) rectangle (2.5cm, 2.5cm);
  \draw[pattern=north west lines, pattern color=gray] (0, 0) circle (2.5cm);
\end{scope}

\begin{scope}
  \clip[rotate=80] ({(cos(70) * 2.5cm}, -2.5cm) rectangle (2.5cm, 2.5cm);
  \draw[pattern=north east lines, pattern color=gray] (0, 0) circle (2.5cm);
\end{scope}

\draw[black,thick] ({2.5cm * cos(60)}, {2.5cm * sin(60)}) -- ({2.5cm * cos(300)},{2.5cm * sin(300)});
\draw[black,thick] ({2.5cm * cos(150)}, {2.5cm * sin(150)}) -- ({2.5cm * cos(10)}, {2.5cm * sin(10)});

\filldraw (0,0) circle (2pt);
\node at (-0.2cm, -0.2cm) {\large $\vc{0}$};

\draw[black,dashed] (0,0) -- ({2.5cm * cos(0)}, {2.5cm * sin(0)});
\draw[black,dashed] (0,0) -- ({2.5cm * cos(80)}, {2.5cm * sin(80)});
\node at (2.8cm, 0cm) {\large $\vc{q}$};
\node at ({2.8cm * cos(80)},{2.8cm * sin(80)}) {\large $\vc{p}$};
\node[black,thick] at (2.5cm,0) {\textbullet};
\node[black,thick] at ({2.5cm * cos(80)},{2.5cm * sin(80)}) {\textbullet};

\draw[<->] (0, -0.2) -- ({2.5cm * cos(60)}, -0.2);
\node at ({1.25cm * cos(60)}, -0.45) {\large $\alphaq$};
\draw[<->] (-0.18cm, 0.04cm) -- ({cos(70) * 2.5cm * cos(80) - 0.18cm},{cos(70) * 2.5cm * sin(80) + 0.04cm});
\node at (-0.5cm, 0.5cm) {\large $\alphau$};

\draw (0.5cm,0cm) arc (0:80:0.5cm);
\node at (0.6cm, 0.4cm) {\large $\theta$};

\draw[black,thick] (0,0) circle (2.5cm);

\draw[pattern=north east lines, pattern color=gray] (4.0cm, 1.25cm) rectangle (5.3cm, 2.25cm);
\node at (7.6cm, 2cm) {$\Pr_{\vc{u} \sim \cS^{d-1}}[\vc{u} \in \text{Update}(\vc{p})]$};
\node at (6.65cm, 1.5cm) {$\propto \cC(\alphau)$};

\draw[pattern=north east lines, pattern color=gray] (4.0cm, -0.5cm) rectangle (5.3cm, 0.5cm);
\draw[pattern=north west lines, pattern color=gray] (4.0cm, -0.5cm) rectangle (5.3cm, 0.5cm);
\node at (8.5cm, 0.25cm) {$\Pr_{\vc{u} \sim \cS^{d-1}}[\vc{u} \in \text{Update}(\vc{p}) \cap \text{Query}(\vc{q})]$};
\node at (7.25cm, -0.25cm) {$\propto \cW(\alphaq, \alphau, \theta)$};

\draw[pattern=north west lines, pattern color=gray] (4.0cm, -2.25cm) rectangle (5.3cm, -1.25cm);
\node at (7.5cm, -1.5cm) {$\Pr_{\vc{u} \sim \cS^{d-1}}[\vc{u} \in \text{Query}(\vc{q})]$};
\node at (6.65cm, -2cm) {$\propto \cC(\alphaq)$};

\end{tikzpicture}
\caption{The geometry of spherical filters. A vector $\vc{p}$ is inserted into a filter $\vc{u}$ with probability proportional to $\cC(\alphau)$, over the randomness of sampling $\vc{u}$ at random from $\cS^{d-1}$. A filter $\vc{u}$ is queried for near neighbors for $\vc{q}$ with probability $\cC(\alphaq)$. A vector $\vc{p}$ at angle $\theta$ from $\vc{q}$ is found as a candidate nearest neighbor in one of the filters with probability proportional to $\cW(\alphaq, \alphau, \theta)$. \label{fig:geometry}}
\end{figure*}

\subparagraph*{Choosing parameters.} To guarantee that nearby vectors at angle $\theta$ are found through a collision, we set $f \sim 1/\cW(\alphaq, \alphau, \theta)$ (up to polynomial terms). What remains is choosing $\alphaq$ and $\alphau$ to obtain a suitable tradeoff between the query and update complexities. Observe that $\alphaq > \alphau$ means that queries return fewer filters than the updates, and so queries are fast but the data structure will use more space. For $\alphaq < \alphau$ we query more filters at a higher query cost, but saving on the space complexity and the update costs.

To balance the query costs of (1) finding the $f \cdot \cC(\alphaq)$ relevant filters, and (2) going through $f \cdot n \cdot \cC(\alphaq) \cdot \cC(\alphau)$ false positives, we choose $\alpha_u$ such that $n \cdot \cC(\alphau) = 1$, i.e.\ we set $\alphau = \sqrt{1 - n^{-2/d}}$. We further set $\alphaq = \beta \alphau$, so that only $\beta$ remains to be chosen, and determines the query/update tradeoff: $\beta < 1$ leads to a better space complexity, and $\beta > 1$ leads to a better (query) time complexity. For $\beta = 1$ we get the results from~\cite{becker16lsf}.

Next, observe that for $\beta = \cos \theta$ the space complexity scales as $n \cdot f \cdot \cC(\alphau) \sim n$, which is the minimum space complexity one can hope for -- further decreasing $\beta$ would only lead to larger query and update exponents. Similarly, in the limit of $\beta \to 1 / \cos \theta$ we obtain the best possible query complexities (subpolynomial in $n$), and further increasing $\beta$ would only lead to worse time and space complexities. This means the optimal parameter range for $\beta$ is $[\cos \theta, 1/\cos \theta]$ and for these parameters we obtain the following result.
\begin{theorem} [Near neighbor tradeoffs]\label{thm:maindense}
Let $\theta \in (0, \frac{1}{2} \pi)$ and let $\beta \in [\cos \theta, 1 / \cos \theta)$. Then using spherical locality-sensitive filters with $\alphau = \sqrt{1 - n^{-2/d}}$ and $\alphaq = \beta \alpha_u$, we can solve the $\theta$-near neighbor problem with update and query exponents given by: 
\begin{align}
\rhou &= \log \left[1 - \left(1 - n^{-2/d}\right) \frac{1 + \beta^2 - 2 \beta \cos \theta}{\sin^2 \theta}\right] / \log(n^{-2/d}) - 1, \\
\rhoq &= \log \left[1 - \left(1 - n^{-2/d}\right) \frac{1 + \beta^2 - 2 \beta \cos \theta}{\sin^2 \theta}\right] / \log(n^{-2/d}) \\
& \qquad \qquad - \log \left[1 - \left(1 - n^{-2/d}\right) \beta^2 \right] / \log(n^{-2/d}).
\label{eq:main2}
\end{align}
This data structure requires $n^{1 + \rhou + o(1)}$ memory, can be initialized in time $n^{1 + \rhou + o(1)}$, allows for updates in time $n^{\rhou + o(1)}$, and answers queries in time $n^{\rhoq + o(1)}$.
\end{theorem}
In the limit of $n^{1/d} = 1 + \eps$ with $\eps \to 0$, corresponding to random data sets in the sparse regime, a Taylor expansion around $\eps = 0$ of $\rhoq$ and $\rhou$ yields:
\begin{align}
\rhou = \frac{(\beta - \cos \theta)^2}{\sin^2 \theta} + o(1), \qquad \rhoq = \frac{(1 - \beta \cos \theta)^2}{\sin^2 \theta} + o(1).
\end{align} 
This leads to the concise defining equation $\sqrt{\rhoq} + \cos \theta \sqrt{\rhou} = \sin \theta$ for the sparse regime, which is equivalent to~\cite[Equation (1)]{andoni17} after substituting $\cos \theta = 1 - 1/c^2$. This also clearly shows how to set $\beta$ to minimize $\rhou$ or $\rhoq$ in the sparse regime.


\section{Finding short tuples with near neighbor searching} 
\label{sec:tuple}

\subparagraph*{Problem description.} In this section we will consider the following problem, which is roughly equivalent to the approximate $k$-list problem defined in~\cite[Definition 1]{herold17}.
\begin{definition}[Tuples on the sphere]
Let $k \geq 2$ and let $L_k$ consist of $n_k$ points drawn uniformly at random from $\cS^{d-1}$. Find most $k$-tuples $(\vc{x}_1, \dots, \vc{x}_k) \in L_k^k$ with $\|\sum_i \vc{x}_i\| \leq 1$.
\end{definition}
Solving this problem allows us to solve SVP with tuple lattice sieving with similar space and time complexities, under the heuristic assumption that throughout the execution of tuple lattice sieving, normalized lattice vectors are uniformly distributed on the sphere. For further details on this heuristic assumption, see e.g.\ \cite{nguyen08, laarhoven15crypto, becker16lsf}. In the application of lattice sieving, one would (1) generate a long list of long lattice vectors; (2) recursively apply a tuple sieve to the list several times; and (3) find a shortest vector in the last non-empty list. 

\subparagraph*{Algorithm description.} As described in the preliminaries, Herold--Kirshanova~\cite{herold17} showed that the global configuration search on $k$-tuples can be efficiently reduced to a local configuration search on pairs of vectors. This directly suggests an approach where (after perhaps initializing some near neighbor data structures) we go through all vectors $\vc{x}_k \in L = L_k$ in the list one by one, and for each $\vc{x}_k$ we find good $k$-tuples that can be formed with $\vc{x}_k$: (1) we first find all vectors in the list satisfying the local configuration property with $\vc{x}_k$, and (2) we then proceed in this smaller list $L_{k-1}$ to find good $(k-1)$-tuples that can be combined with $\vc{x}_k$ to form a short $k$-tuple. An outline of this approach is given in Algorithm~\ref{alg:tuple}.

\begin{algorithm}[H]
\caption{$\textsc{TupleSieve}(k, L_k)$}
\label{alg:tuple}
\begin{algorithmic}[1]
\Require Tuple size $k \geq 2$, input list $L_k \subset \cS^{d-1}$ uniformly distributed on a unit sphere
\Ensure Returns (almost) all $k$-tuples $(\vc{x}_1, \dots, \vc{x}_k) \in L_k^k$ with $\|\sum_i \vc{x}_i\| < 1$ 
\State $\theta_k \gets \arccos(-\frac{1}{k})$ \Comment{\textit{$\theta_k$: target local configuration}}
\State Initialize a $\theta_k$-NN data structure $\cD_k$ \label{lin:build1}
\For{\textbf{each} $\vc{x}_k \in L_k$}
	\State $\textsc{Insert}(\cD_k, \vc{x}_k)$
\EndFor \label{lin:build2}
\State $S_k \gets \emptyset$ \Comment{\textit{$S_k$: set of all $k$-tuple solutions}}
\For{\textbf{each} $\vc{x}_k \in L_k$}
	\State $L_{k-1} \gets \textsc{Query}(\cD_k, \vc{x}_k)$ \Comment{\textit{$L_{k-1}$: all $\theta_k$-near neighbors to $\vc{x}_k$}} \label{lin:query}
	\If{$k > 2$}
		\State $L_{k-1}' \gets \textsc{Transform}(k, L_{k-1}, \vc{x}_k)$	\label{lin:1}\Comment{\textsc{Transform}: \textit{see text below}} 
		\State $S_{k-1}' \gets \textsc{TupleSieve}(k-1, L_{k-1}')$ \label{lin:rec}
		\State $S_{k-1} \gets \textsc{Transform}^{-1}(k, S_{k-1}', \vc{x}_k)$ \label{lin:2} \Comment{\textit{$S_{k-1}$: tuples combinable with $\vc{x}_k$}}
	\EndIf
	\State $S_k \gets S_k \cup (S_{k-1} \times \{\vc{x}_k\})$
\EndFor
\State \Return $S_k$
\end{algorithmic}
\end{algorithm}

\subparagraph*{Transform.} Ideally, one might hope that Algorithm~\ref{alg:tuple} cleanly solves the problem recursively without requiring the transforms in Lines~\ref{lin:1} and \ref{lin:2}. However, since all vectors in $L_{k-1}$ have inner product approximately $-\frac{1}{k}$ with $\vc{x}_k$, the list $L_{k-1}$ is clearly not distributed uniformly on the unit sphere. To solve this problem, observe that if $L_k$ is uniformly distributed on $\cS^{d-1}$, then $L_{k-1}$ is essentially uniformly distributed on $\cS^{d-1} \cap \cP_{\vc{x}_k, \frac{-1}{k}}$, with $\cP_{\vc{u}, \vc{\alpha}}$ denoting the hyperplane in $\mathbb{R}^d$ with defining equation $\ip{\vc{x}}{\vc{u}} = \alpha$. This set is isomorphic to $\cS^{d-2}$, and via a simple transformation of $L_{k-1}$ (e.g.\ applying an orthogonal rotation which maps $\vc{e}_k$ onto $\vc{x}_k$, and then projecting onto the first $k-1$ coordinates) we can map $L_{k-1}$ to a set $L_{k-1}'$ which is again uniformly distributed on a unit sphere, albeit of one dimension less. 

\subparagraph{Transforming local configurations.} To understand the effects of the transform, and to understand how the local configurations of Lemma~\ref{lem:conf} change after the transform, suppose w.l.o.g.\ that $\vc{x}_k = \vc{e}_k$. Then $L_{k-1}$ contains vectors uniformly distributed on $\{\vc{x} \in \cS^{d-1}: x_k = \frac{-1}{k}\}$. After the projection eliminates the last coordinate, we are left with $(k-1)$-dimensional vectors of norm $\sqrt{k^2 - 1}/k$, so the transform scales the entire data set by $k/\sqrt{k^2 - 1}$ to obtain unit vectors. The transform $\cT: \cS^{d-1} \to \cS^{d-2}$ can thus be modeled as:
\begin{align}
\cT (u_1, \dots, u_k) = \frac{k}{\sqrt{k^2 - 1}} \, (u_1, \dots, u_{k-1}).
\end{align}
From Lemma~\ref{lem:conf}, in the next recursive step we are again looking for vectors $\vc{y}, \vc{z}$ with pairwise inner product $-\frac{1}{k}$. Let $\vc{y}, \vc{z} \in L_k$, satisfying the local configuration property with $\vc{x}_k$, and let $\vc{y}' = \cT \vc{y}$ and $\vc{z}' = \cT \vc{z}$ denote their images after the transform. The condition $\ip{\vc{y}}{\vc{z}} \approx -\frac{1}{k}$ then translates to a condition on $\ip{\vc{y}'}{\vc{z}'}$ as:
\begin{align}
\ip{\vc{y}'}{\vc{z}'} = \frac{k^2}{k^2 - 1} \Big(\ip{\vc{y}}{\vc{z}} - y_k z_k\Big) = \frac{k^2}{k^2 - 1} \left(\frac{-1}{k} - \frac{1}{k^2}\right) = \frac{-1}{k-1}
\end{align} 
In other words, after the transform, we have a set of uniformly distributed vectors on a unit sphere of one dimension less, and the target local configuration (inner product) has changed from $-\frac{-1}{k}$ for $k$-tuples to $\frac{-1}{k-1}$ for $(k-1)$-tuples. After searching for pairwise inner products $\frac{-1}{k}$ in the outermost loop, we are therefore searching for pairwise inner products $\frac{-1}{k-1}$ in the next loop, $\frac{-1}{k-2}$ in the next loop etc., and inner product $\frac{-1}{2}$ in the innermost loop. 

Algorithm~\ref{alg:tuple} describes this approach recursively, with the transform as outlined above and $\theta_k$ denoting the target configurations of inner product $\frac{-1}{k}$ when looking for $k$-tuples.

\subparagraph*{List sizes at given levels.} For analyzing the complexities of this algorithm, we need to know how many vectors are left at each recursive call of the sieve, i.e.\ how big $L_{k-1}$ is relative to $L_k$. As the target angle is $\theta_k = \arccos(\frac{-1}{k})$, a fraction $\cC(\frac{-1}{k}) = \cC(\frac{1}{k})$ of all points in $L_k$ are expected to be in $L_{k-1}$ for any choice of $\vc{x}_k$. After $k-i$ vectors $\vc{x}_k, \vc{x}_{k-1}, \dots, \vc{x}_{i+1}$ have been fixed, we are expected to be left with a list $L_i$ containing a fraction $\prod_{j=i+1}^k \cC(\frac{1}{j})$ of all vectors in $L_k$. As the product telescopes, for constant $k$ this can be simplified to:
\begin{align}
\frac{|L_i|}{|L_k|} = \prod_{j=i+1}^k \cC(\tfrac{1}{j}) = \left(\prod_{j=i+1}^k \frac{(j - 1)(j + 1)}{j^2}\right)^{d/2 + o(d)} = \left(\frac{(k + 1) \cdot i}{k \cdot (i + 1)}\right)^{d/2 + o(d)}.
\end{align}
In case $|L_k|$ is chosen as in Lemma~\ref{lem:ls}, we find that for the $k$-tuple sieve, after $k-i$ vectors have been fixed, we are left with a list of vectors satisfying the pairwise inner product constraint with these $k-i$ vectors of expected size as follows, similar to \cite[Equation (16)]{herold17}.
\begin{align}
n_i := |L_i| = \left(\frac{(k + 1) \cdot i}{k \cdot (i + 1)}\right)^{d/2 + o(d)} \cdot \left(\frac{k^{k/(k-1)}}{k+1}\right)^{d/2 + o(d)} = \left(\frac{k^{1/(k-1)} \cdot i}{i + 1}\right)^{d/2 + o(d)}.
\end{align}
For large $k$ and small $i$, the expected size of this list may become smaller than $1$: in most cases there are no $k$-tuples that can be constructed using this choice of $\vc{x}_k, \vc{x}_{k-1}, \dots, \vc{x}_{i+1}$, and in some cases there are only a few solutions. In the tree of choices for the tuple vectors (similar to the enumeration tree in enumeration-based SVP solvers), this means that in the lower levels of the tree many branches will lead to dead ends, and the number of leaves is significantly smaller than the maximum width of the tree.

\subparagraph*{Tree widths at given levels.} Finally, let us also consider the width of the enumeration-like tree of vector combinations at different levels $i$, which will be relevant for computing the time complexity of the tuple sieve. Up to a factor $(k - i)! = O(1)$ due to double counting, there are $\prod_{j=i+1}^k |L_j|$ different choices for the first $k - i$ vectors in the tuple, and this product can be somewhat simplified to:
\begin{align}
w_i := \prod_{j=i+1}^k |L_j| 
= \left(k^{(k-i)/(k-1)} \cdot \frac{i + 1}{k + 1}\right)^{d/2 + o(d)}.
\end{align}
For $i = 0$, so that all $k$ vectors have been chosen, we are left with a total of $w_0 = \prod_{j=1}^k |L_j| \sim |L_k|$ vectors, which exactly matches the condition that the output list of good $k$-tuples should be as large as the input list. The number $w_i$ can also be interpreted to the number of calls to \textsc{TupleSieve$(i, L_i)$} for some list $L_i$.

\subparagraph*{Results.} With the above expressions for (the products of) the list sizes at hand, we can now provide explicit asymptotics for the time and space complexities, in terms of the near neighbor exponents $\rhoq^{(i)}$ and $\rhou^{(i)}$ at each level in the tree.
\begin{theorem}[General time and space complexities] \label{thm:tuple}
Let $(\rhou^{(i)}, \rhoq^{(i)})$ denote the update and query exponents for the near neighbor data structure $\cD^{(i)}$ at level $i$. Then we can solve tuples of the sphere with $L_k$ as in Lemma~\ref{lem:ls} in time $\mathrm{T}$ and space $\mathrm{S}$, with:
\begin{align}
\mathrm{S} &= \max \Big\{n_i^{1 + \rhou^{(i)}}: i = 1, \dots, k\Big\}, \\
\mathrm{T} &= \max \Big\{w_i \cdot \max\Big\{1, n_i^{1 + \rhou^{(i)}}, n_i^{1 + \rhoq^{(i)}}\Big\}: i = 1, \dots, k\Big\}.
\end{align}
\end{theorem}
Here the three terms in the time complexity correspond to the number of recursive calls to the \textsc{TupleSieve}, the creation of the data structure in Lines~\ref{lin:build1}--\ref{lin:build2}, and the queries to the data structure in Line~\ref{lin:query}. As the tuple sieve at level $i$ is essentially called $w_i$ times with similar parameters each time, this explains the leading factors $w_i$ in the time complexities. The space complexity is simply the maximum over the space complexities of the near neighbor data structures at each level. Note that $\rhou^{(i)} > 0$ implies that the data structures require space at least linear in the list sizes $n_i$.

\subparagraph*{Choosing parameters.} Assuming that there is a strict tradeoff between $\rhoq$ and $\rhou$ at each level, choosing parameters can now be done as follows, regardless of the near neighbor method used. Suppose we want to find the best overall time complexity for a given overall space complexity $\mathrm{S}$. If choosing $\rhoq^{(i)} = \rhou^{(i)}$ leads to exponents with $n_i^{1 + \rhou^{(i)}} \leq \mathrm{S}$, then this is optimal - the time complexity symmetrically depends on $\rhoq^{(i)}$ and $\rhou^{(i)}$, so the best choice is to balance them. If a balanced choice leads to $n_i^{1 + \rhou^{(i)}} \gg \mathrm{S}$, then we cannot use these parameters as it will take too much space, and so we need to decrease $\rhou^{(i)}$ until we reach $n_i^{1 + \rhou^{(i)}} \approx \mathrm{S}$. This choice $\rhou^{(i)}$ then leads to the smallest allowed value $\rhoq^{(i)}$ that can be used. Note that we can always find such a value $\rhou^{(i)}$ with $n_i^{1 + \rhou^{(i)}} \leq \mathrm{S}$ unless $\mathrm{S} \ll n_i \leq n_k$, in which case the available memory is less than the size of the input/output lists.


\section{Combining tuple lattice sieving with spherical filters}
\label{sec:tuple2}

Let us finally describe the results that can be obtained by applying the spherical filters of Section~\ref{sec:dense} to the tuple sieving approach of Section~\ref{sec:tuple}. As spherical filters have so far led to the best asymptotic exponents for sieving with pairwise reductions~\cite{becker16lsf}, one might expect that this also leads to the best exponents for tuple lattice sieving. However, no matching lower bounds for the high-density regime of near neighbor searching are known, and in practice other methods such as hyperplane/hypercube LSH~\cite{charikar02, laarhoven15crypto, laarhoven17hypercube, mariano15, mariano17} or cross-polytope LSH~\cite{terasawa07, andoni15cp, becker16cp} may lead to better results. An experimental evaluation of different near neighbor methods for tuple lattice sieving is left for future work.

\subparagraph*{Triple lattice sieving.} For $k = 3$, the input list has size $n_3 = 2^{0.1887d + o(d)}$, while the filtered lists have expected size $n_2 = 2^{0.1037d + o(d)}$ and $n_1 = 2^{-0.1037d + o(d)}$. Focusing on the near-linear space regime with memory bounded by $2^{0.1887d + o(d)}$, we get the following parameters for the different loops:
\begin{itemize}
\item For the outermost loop, we need to set $\rhou^{(3)} = 0$. This corresponds to setting $\beta^{(3)} = \cos \theta_3 = \frac{-1}{3}$ in Theorem~\ref{thm:maindense}, leading to $\rhoq^{(3)} = 0.9010$.
\item For the search over pairs, we need to set $\rhou^{(2)} \leq 0.8188$ due to the memory restriction. As balancing the query and update exponents with $\beta^{(2)} = 1$ leads to $\rhoq^{(2)} = \rhou^{(2)} = 0.3681 < 0.8188$, this choice leads to the best time complexity for the inner loop.
\end{itemize}
This together implies that the space is bounded by $2^{0.1887d + o(d)}$, and (1) the outer loop takes time $2^{0.1887d + o(d)}$ to initialize/build the data structure $\cD_3$; (2) the outer loop takes time $2^{0.3588d + o(d)}$ to query the data structure for each $\vc{x}_3$; and (3) the inner loops take total time $2^{0.3307d + o(d)}$ to find suitable triples. Overall the time complexity is therefore $2^{0.3588d + o(d)}$, improving upon the $2^{0.3717d + o(d)}$ of~\cite{herold17}.

\subparagraph*{Larger tuple sizes.} As explained at the end of Section~\ref{sec:tuple}, optimizing parameters is straightforward when given an explicit description of the near neighbor tradeoffs for arbitrary list sizes and target angles. For arbitrary tuple sizes, with memory limited to near-linear in the list size, Table~\ref{tab:1} lists the resulting optimized time complexity exponents when using spherical locality-sensitive filters. All these time complexities are improvements over~\cite{bai16, herold17}.  

\begin{table}[!b]
\begin{center}
\begin{tabular}{p{2.4cm}p{0.8cm}p{0.8cm}p{0.8cm}p{0.8cm}p{0.8cm}p{0.8cm}p{0.8cm}p{0.8cm}p{0.8cm}} \toprule
Tuple size ($k$) & $2$ & $3$ & $4$ & $5$ & $6$ & $7$ & $8$ & $9$ & $10$ \\ \midrule
$\textrm{Space complexity}$ & $0.2075$ & $0.1887$ & $0.1724$ & $0.1587$ & $0.1473$ & $0.1376$ & $0.1293$ & $0.1221$ & $0.1158$ 
\\ 
Time complexity & $0.3685$ & $0.3588$ & $0.3766$ & $0.4159$ & $0.4497$ & $0.4834$ & $0.5205$ & $0.5510$ & $0.5767$ 
\\ \bottomrule
\end{tabular}
\vspace{0.1cm}
\caption{Leading constants in the exponent of the time and space complexities of tuple lattice sieving. For instance, the optimized quadruple sieve runs in time $2^{0.3766d + o(d)}$ and space $2^{0.1724d + o(d)}$, improving upon the previous best time complexity of $2^{0.4080d + o(d)}$ of~\cite{herold17}.} \label{tab:1}
\end{center}
\end{table}

\subparagraph*{Complete tradeoff spectrum.} We can also easily obtain further tradeoffs between the time and space complexities, by setting the maximum available memory to be (slightly) larger than the list size. It turns out that as the tuple size increases, this quickly does not lead to any useful results: increasing the available memory hardly leads to an improvement in the time complexity, if any. For small tuple sizes however we do obtain a significant improvement, e.g.\ for $k = 3$ the best time complexity is obtained when the space is $2^{0.2108d + o(d)}$, with time complexity $2^{0.3307d + o(d)}$. When the space is equal to the list size for classical sieving methods ($2^{0.2075d + o(d)}$), our best time complexity with spherical filtering is $2^{0.3317d + o(d)}$.

\begin{figure*}[!t]
\centering
\includegraphics[width=\linewidth]{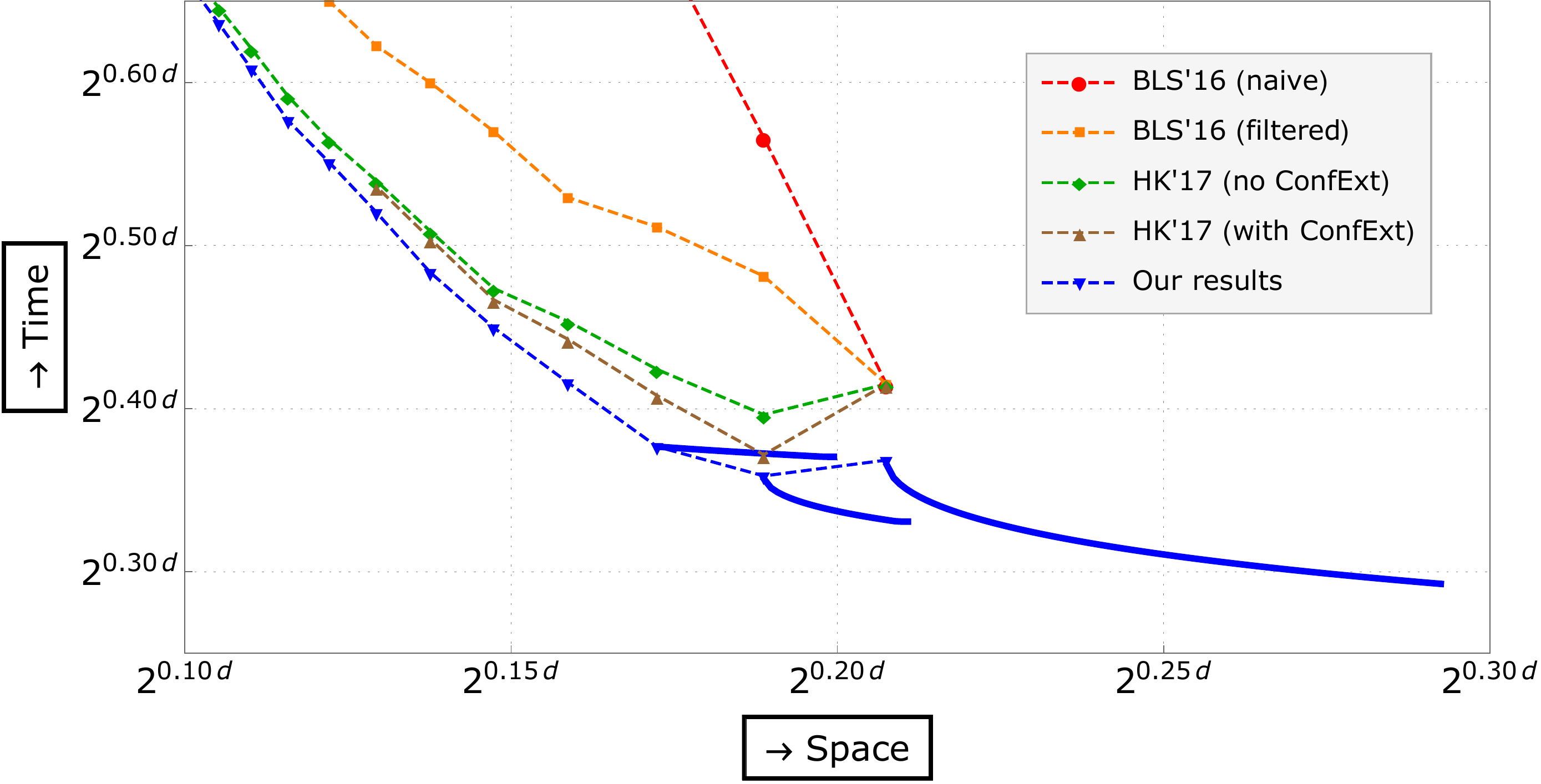}
\caption{Tradeoffs between the time and space complexities for all tuple sizes. Dashed lines are only present to highlight points corresponding to the same algorithm, whereas thick curves represent actual achievable space-time tradeoffs using spherical filtering.\label{fig:tradeoff}}
\end{figure*}

Figure~\ref{fig:tradeoff} illustrates the non-trivial tradeoffs that can be obtained for tuples sizes $k = 2, 3, 4$. These tradeoffs, together with the optimized tuple sieving complexities of Table~\ref{tab:1} (blue), are compared against the previous best tuple sieving complexities of~\cite{bai16, herold17}. The tradeoff for $k = 2$ overlaps with the best space-time tradeoff of~\cite{becker16lsf} for sieving with pairwise reductions. Note that the green points in Figure~\ref{fig:tradeoff}, corresponding to results from~\cite{herold17} without using Configuration Extension, are equivalent to the complexities of tuple sieving \textit{with} an optimized configuration search, but textit{without} using any near neighbor techniques (i.e.\ setting $\rhou^{(i)} = 0$ and $\rhoq^{(i)} = 1$ in Theorem~\ref{thm:tuple}).

\subparagraph*{Large-$k$ asymptotics.} As described in Theorem~\ref{thm:tuple}, the time complexity for tuple sieving is lower-bounded by the largest $w_i$, and upper-bounded by the largest value of $w_i \cdot n_i$ (which can be obtained with a linear search, with exponents $\rhou^{(i)} = 0$ and $\rhoq^{(i)} = 1$). Since $n_i \leq n_k$ approaches $0$ for large $k$, and $w_i$ grows with $k$, it is clear that the effects of near neighbor searching become less and less as $k$ increases. This can also be observed in Figure~\ref{fig:tradeoff}: as $k$ increases and the space complexity decreases, the blue dashed ``curve'' using near neighbor searching approaches the green ``curve'' based on an optimal configuration search, without using any near neighbor techniques.

\subparagraph*{Practical estimates.} To predict the practicability of our tuple sieving approach (in particular triple sieving), observe that while the time complexities are worse than the best time-optimized double sieve (time and space $2^{0.2925d + o(d)}$~\cite{becker16lsf}), the space complexity is significantly better. When using $2^{0.2075d + o(d)}$ memory for a triple sieve, we save a factor $2^{0.0850d + o(d)}$ on the space complexity, at the cost of $2^{0.0392d + o(d)}$ more time. Since sieving with near neighbor searching is memory-bound~\cite{mariano16pdp, mariano17}, the improvement in the memory complexity will likely lead to a better time complexity as well. If we further compare this triple sieve with the practical HashSieve~\cite{laarhoven15crypto, mariano15} with time and space $2^{0.3366d + o(d)}$, which was previously used to solve SVP in dimension 107~\cite{svp}, we see that asymptotically we save even more on the time complexity, while saving on the space complexity as well. If the hidden order terms of triple sieving with near neighbor searching are not too large, one should also be able to solve SVP in dimension $100$ or higher with a triple sieve without too much effort. 

Open questions for future work are whether the hidden order terms of triple sieving are sufficiently small, and whether spherical LSF is the right choice in practice. Especially for the inner loop, searching for near neighbors in a list of size $n_2 < 2^{0.11d + o(d)}$, the list will be rather short and a more basic approach like hyperplane/hypercube LSH~\cite{charikar02, laarhoven17hypercube}, cross-polytope LSH~\cite{terasawa07, andoni15cp}, or even a linear search may lead to better complexities in practice.

\subparagraph*{The Nguyen-Vidick sieve and the GaussSieve.} For the underlying lattice sieve of tuple sieving, there are two different approaches: (1) the AKS-style Nguyen--Vidick sieve~\cite{ajtai01, nguyen08}, where after generating a long initial list, the list size is reduced each step; and (2) Micciancio--Voulgaris' GaussSieve~\cite{micciancio10b}, where one starts with an initially empty list, and only adds new vectors when no more progress can be made. Whereas numerous experimental works have proven the practicality of the GaussSieve~\cite{milde11, schneider11, schneider13, ishiguro14, mariano14, mariano14b, fitzpatrick14, bos16, yang17}, no competitive implementations of the Nguyen--Vidick sieve approach exist to date. Ideally, tuple sieving should therefore also be based on the GaussSieve to be practical. 

Tuple sieving can be based on either method, including the GaussSieve. To illustrate how to do this with the triple sieve, recall that there is an outer list over triples, and an inner search for pairs, given one vector of the tuple. To implement this method with the GaussSieve, one first initializes a global data structure for the outer searches. Each time a vector is chosen to be reduced against the list, the inner near neighbor data structure is reinitialized for the shorter list, used to find good triples, and erased from memory again. If a vector in the list must be updated, the outer near neighbor data structure needs to be updated as well, which can fortunately be done quite efficiently.

\subparagraph*{Saving space with the Nguyen--Vidick sieve.} For double sieving, an argument can be made for the Nguyen--Vidick sieve when using near neighbor techniques, as one can theoretically save on the space complexity by processing the near neighbor data structure sequentially (for details, see~\cite{becker15nns, laarhoven15crypto}). The same tricks however do not apply to tuple sieving. To see why, note that processing the outer ``hash tables'' sequentially would mean that in the inner loop, we process subsets $L_{1,1}, \dots, L_{1,T} \subset L_1$ sequentially, with $\bigcup_{i=1}^T L_{1,i} = L_1$. For the double sieve, solutions in this inner loop are simply elements from these lists, and the union of the ``$1$-tuple solutions'' of these smaller lists is the same as the solutions of the entire list $L_1$.

The same however does not hold for the outer search in the triple sieve: the union of $2$-tuple solutions in each $L_{2,i}$ is not the same as the $2$-tuple solutions in $L_2$, as the former excludes all cases where the two vectors in this $2$-tuple come from different sets $L_{2,i}, L_{2,j}$ with $i \neq j$. Only the inner loop can therefore (theoretically) be improved to use only linear space, when using the Nguyen--Vidick sieve instead of the GaussSieve. However, for none of the tuple sieves, the inner loop dominates the time complexity, and speeding up the inner loop will not decrease the theoretical time complexity exponent. 


\subsection*{Acknowledgments}

The author thanks Gottfried Herold and Elena Kirshanova for discussions regarding tuple lattice sieving, and for providing their optimal time complexities with ConfExt for tuple sizes $k > 3$. The author further thanks Bryan Smeets for his contributions in May 2016. The author is supported by the SNSF ERC Transfer Grant CRETP2-166734 FELICITY.


\newpage
\bibliographystyle{plain}

\end{document}